\documentclass[aps,prl,twocolumn,groupedaddress]{revtex4}
\usepackage{graphicx}
\usepackage{delarray}%
\usepackage{amsmath}%
\usepackage{amsfonts}%
\usepackage{amssymb}

\begin{document}
\title{Spin-3/2  random quantum antiferromagnetic chains}
\author{A. \surname{Saguia}}
\affiliation{Centro Brasileiro de Pesquisas F\'{\i}sicas \\
Rua Dr. Xavier Sigaud 150 - Urca,  Rio de Janeiro, 22290-180, RJ - Brazil.}
\author{B. \surname{Boechat}}
\author{M. A. \surname{Continentino} }
\email{mucio@if.uff.br}
\affiliation{Departamento de F\'{\i}sica - Universidade Federal Fluminense\\
Av. Litor\^anea s/n,  Niter\'oi, 24210-340, RJ - Brazil}
\date{\today}

\begin{abstract}
We use a modified perturbative renormalization group approach to study the random quantum antiferromagnetic
spin-3/2 chain. We find that in the case of rectangular distributions there is a quantum Griffiths phase and we
obtain the dynamical critical exponent $Z$ as a function of disorder. Only in the case of extreme disorder,
characterized by a power law distribution of exchange couplings, we find evidence that a random singlet phase
could be reached. We discuss the differences between our results and those obtained by other approaches.
\end{abstract}

\maketitle

The study of random antiferromagnetic chains is an important and
actual area in magnetism. Since by now, many of the physical
properties of pure chains are understood, it is natural to include
disorder in these systems and look for the modifications it
introduces. In the case of spin-1/2 random exchange Heisenberg
antiferromagnetic chains (REHAC) a perturbative approach
introduced by Ma,  Dasgupta and Hu (MDH) was very successful
\cite{MDH} to investigate these systems. This approach turns out
to be asymptotically exact and this allowed Fisher \cite{fisher}
to fully characterize the properties of the new disordered phase,
for which, the name {\em random singlet phase} was coined.
Unfortunately when generalized to higher spins this method, in its
simplest version at least, revealed to be ineffective. The reason
is that in the elimination procedure of the strongest bond
$\Omega$, the new interaction between the spins, coupled by
exchanges $J_1$ and $J_2$ to the strongest coupled pair is given
by, $J^{\prime}=(2/3)S(S+1)J_1J_2/\Omega$ \cite{bia}. For $S \ge
1$ the factor $(2/3)S(S+1)>1$ and the problem becomes essentially
non-perturbative for arbitrary distributions of exchange
interactions. Several approaches have been introduced to
circumvent this difficulty
\cite{hyman,jolicoeur,hida1,hida2,bia1,new,bia2,bia4,kedar,saguia,refael,huse,igloi}
which however not always lead to the same unambiguous results. In
this Communication we apply a previous method used to treat the
random spin-1 chain \cite{saguia} for the spin-3/2 REHAC. This
particular chain has been the subject of recent studies
\cite{refael,huse,igloi} and it would be nice to confirm the
results obtained in these works using another approach.

As mentioned previously the method of Ma, Dasgupta and Hu consists
in finding the strongest interaction ($\Omega$) between pairs of
spins in the chain (see Fig.1a) and treating the couplings of this
pair with its neighbors ($J_1$ and $J_2$) as a perturbation. For a
chain of  spins $S=3/2$, after elimination of the strongest
coupled pair, the  new coupling between its neighbors is given by

\begin{equation}
\label{1} J^{\prime}=\frac{5}{2}\frac{J_1 J_2}{\Omega}
\end{equation}

Consider the case $J_1 \ge J_2$.  If $J_1> (2/5) \Omega$ than the new effective interaction $J^{\prime}$ is
necessarily larger than one of those eliminated, in this case,
than $J_2$.

\begin{figure}
 \includegraphics[scale=0.5]{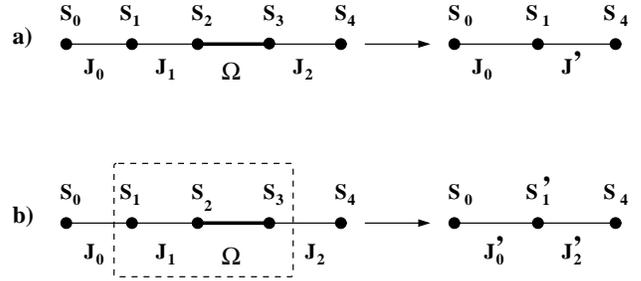}
 \caption{\label{fig1}The two elimination procedures as
 described in the text ($J_1 > J_2$).}
\end{figure}

\begin{figure}
\includegraphics[angle=-90,scale=0.3]{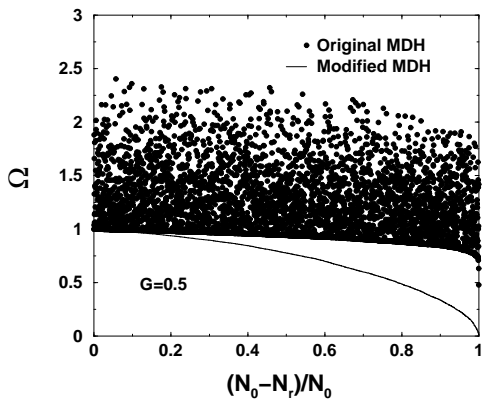}
\caption{\label{fig2} The evolution of the cut-off, for an initial rectangular distribution with $G=0.5$, along
the renormalization process of the spin-$3/2$ REHAC. We show the results for the original MDH and the present
(modified) renormalization group procedures.}
\end{figure}

Our generalization of the $MDH$ method consists in either of the following procedures
shown in Fig.~\ref{fig1}. If
the largest neighboring interaction to $\Omega$, $J_1 < (2/5) \Omega$, then we
eliminate the strongest coupled
pair obtaining an effective interaction between the neighbors to this pair which is given
by Eq.~\ref{1} (see
Fig.~1a). This new effective interaction is always smaller than those eliminated.
Now suppose $J_1 > J_2$ and $J_1
> (2/5) \Omega$. In this case, we consider the {\em trio} of spins $S=3/2$ coupled
by the two strongest interactions of the trio, $J_1$ and $\Omega$
and solve it exactly (see Fig.~1b). The ground state of this trio
of spins $S=3/2$ is a degenerate quadruplet. It will be
substituted by an effective spin$-3/2$ interacting with its
neighbors  through {\em new renormalized} interactions obtained by
degenerate perturbation theory acting on the ground state of the
trio. This procedure which implies diagonalizing the $64X64$
matrix of the trio is carried out {\em analytically}. This is
important for obtaining results on large chains and to deal with
the large numbers of initial configurations that we use. These
procedures guarantee that we always comply with the criterion of
validity of perturbation theory as shown in Fig.~\ref{fig2}. We
have considered initial rectangular distributions,
$P_0(J)=(1/(\Omega-G))\Theta(\Omega-J)\Theta(J-G)$ of interactions
and even for the weak disorder case, with a gap $G$ as large as
$G=0.5$ in the distribution ($\Omega=1$), the method works very
well and never an interaction larger than those eliminated is
generated (see Fig.~\ref{fig2}).

\begin{figure}
\centering
{\includegraphics[angle=-90,scale=0.38]{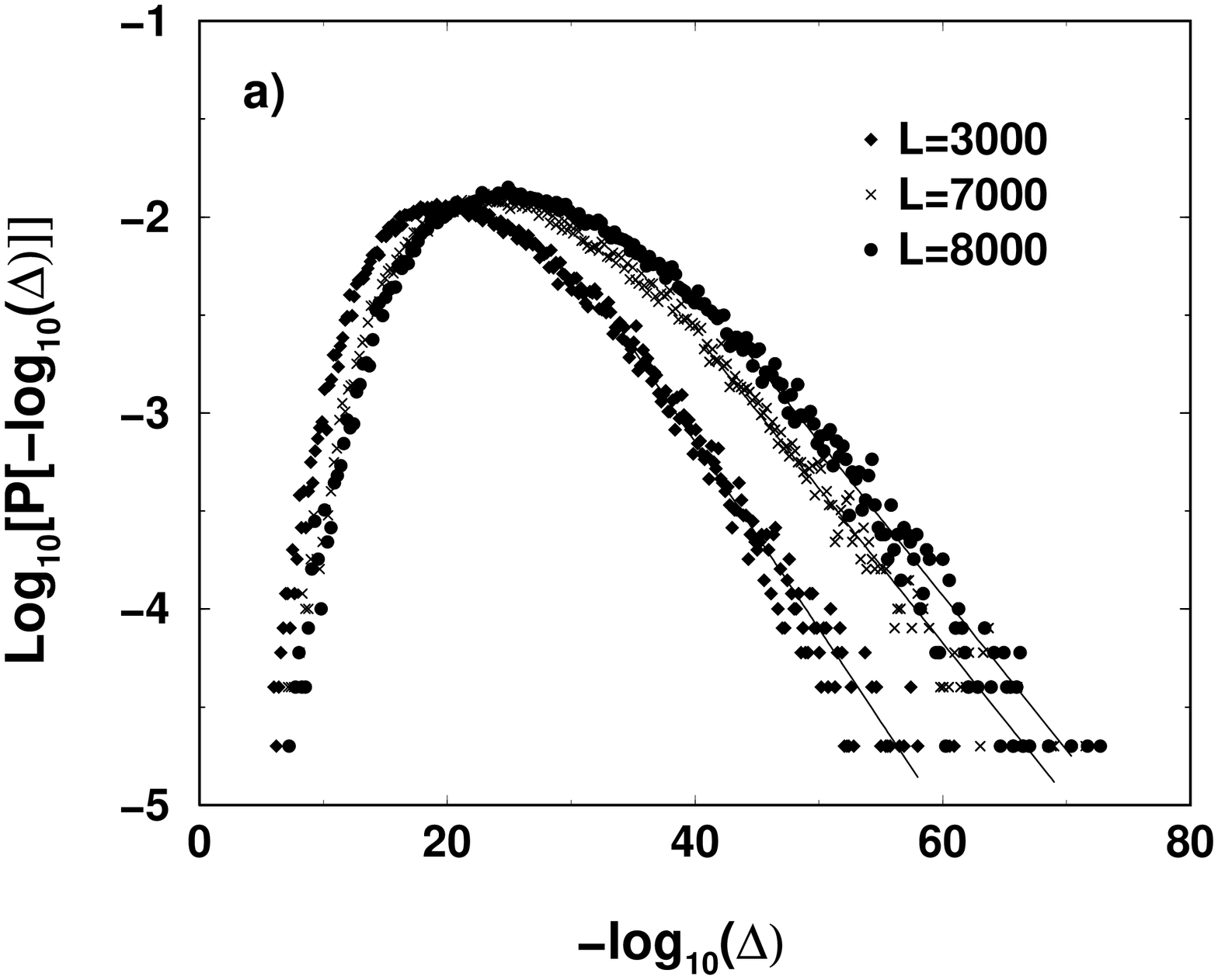}}
{\includegraphics[angle=-90,scale=0.38]{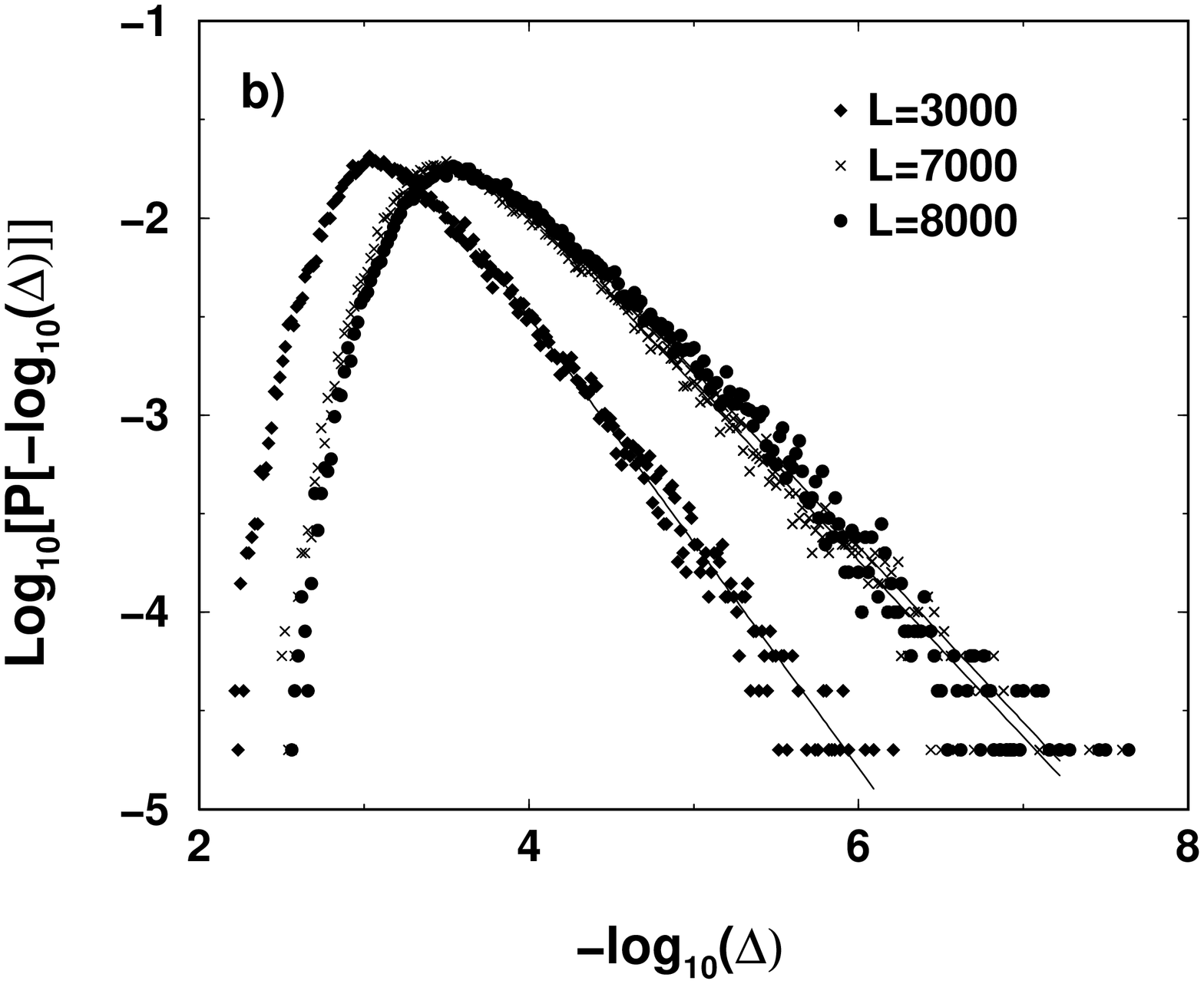}}
{\includegraphics[angle=-90,scale=0.38]{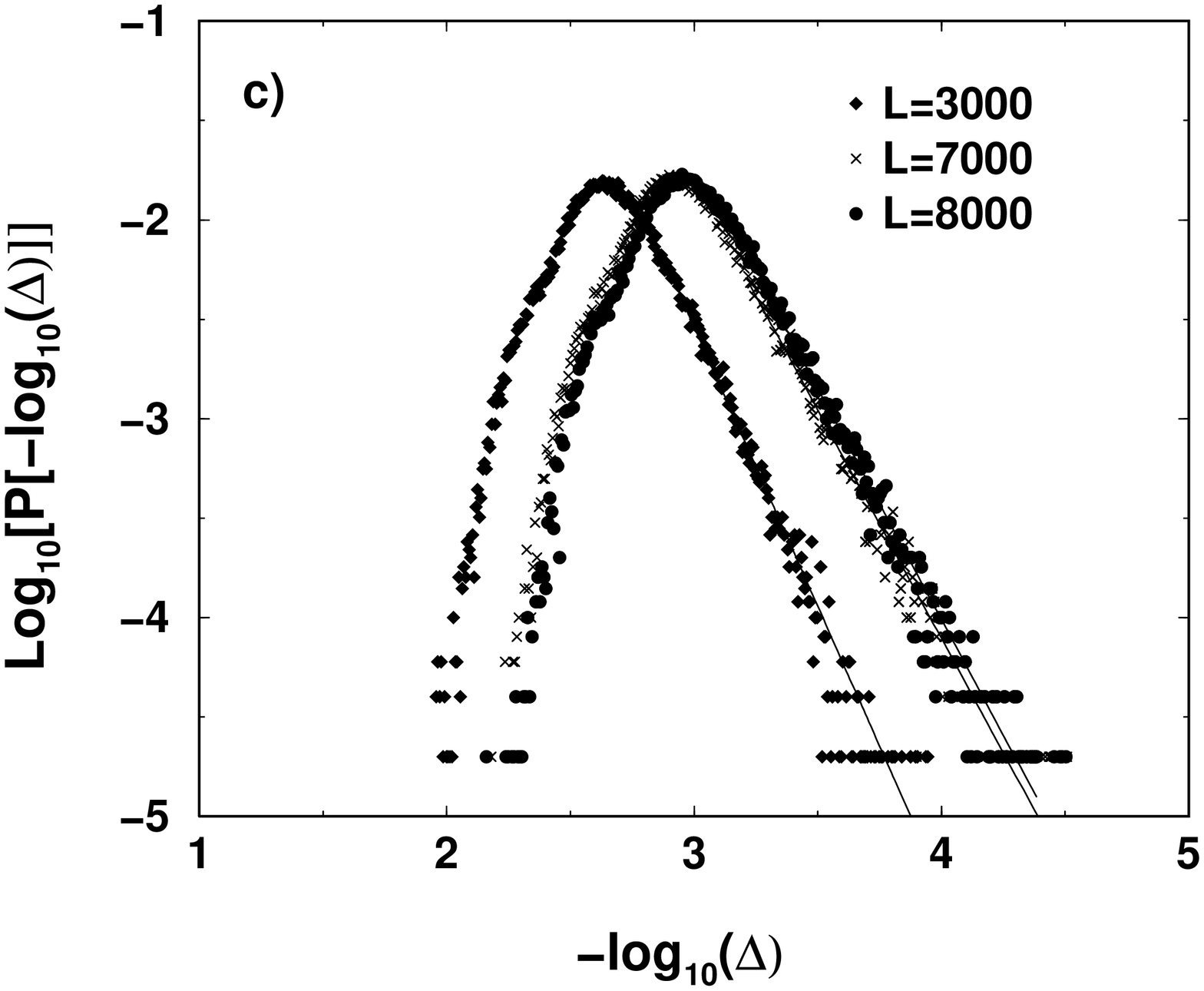}} \caption{ \label{fig3} Probability distributions of {\em first
gap} for initial rectangular distributions of couplings with gaps $G$ and different systems sizes $L$. For clarity
not all values of $L$ are shown. The solid lines represent best fits to the form $\log_{10}[P(-\log_{10}
\Delta)]=A_L-\frac{1}{Z_L} \log_{10} \Delta$. a) $G=0$, $Z_{3000}=10.51$, $Z_{7000}=12.68$, and $Z_{8000}=12.70$.
b) $G=0.12$, $Z_{3000}=0.87$, $Z_{7000}=1.11$, and $Z_{8000}=1.12$. c) $G=0.2$, $Z_{3000}=0.35$, $Z_{7000}=0.43$,
and $Z_{8000}=0.43$.}
\end{figure}

The phase diagram for rectangular original distributions can now
be obtained. For strong disorder, which corresponds to $G=0$, we
find a Griffiths phase with a dynamic exponent $Z \sim 12.7$ as
shown in Fig.~\ref{fig3}. This phase is characterized by {\em
first gap distributions} that saturate at low energies in the
form, $P(-\log \Delta) \sim \Delta^{1/Z}$ for $\Delta \rightarrow
0$~\cite{igloi1,igloi2}. This is obtained starting from a given
configuration of random interactions for a chain of size $L$ and
eliminating the spins, as described above, until a single pair
remains. The interaction between these remaining spins yields the
first gap $\Delta$ for excitation. The dynamic exponent $Z$
relates the scales of length and energy through $\Delta \propto
L^{-Z}$.

In Figure~\ref{fig3} we show the first gap distributions for different degrees of disorder as characterized by
different gaps $G$ in the initial distribution of interactions. For all cases, including that of strong disorder
($G=0$),  we find that the first gap distributions saturate at low energies with the {\em dynamic exponent} $Z$
independent of $L$ for $L$ sufficiently large. We have to consider large chains in order to observe this effect.
We find $Z_{\infty} \sim 0.43$ and $Z_{\infty} \sim 1.12$ for $G=0.2$ and  $G=0.12$, respectively. From these
values of the dynamic exponent we can deduce the existence of a Griffiths phase extending up to $G_c \approx 0.11$
where the dynamic exponent reaches the value $Z=1$. Since, for example, the susceptibility $\chi \propto T^{1-Z}$,
a singular low temperature behavior implies $Z>1$. At $G_c$ there is in fact a significant change in the nature of
the thermodynamic behavior of the system \cite{igloi}. The phase  for $G>G_c$ is one with quasi-long range order,
i.e., with spin correlations decaying algebraically with distance,  similar to the zero temperature phase of the
pure chain \cite{igloi}.

\begin{figure}
\centering
{\includegraphics[angle=-90,scale=0.195]{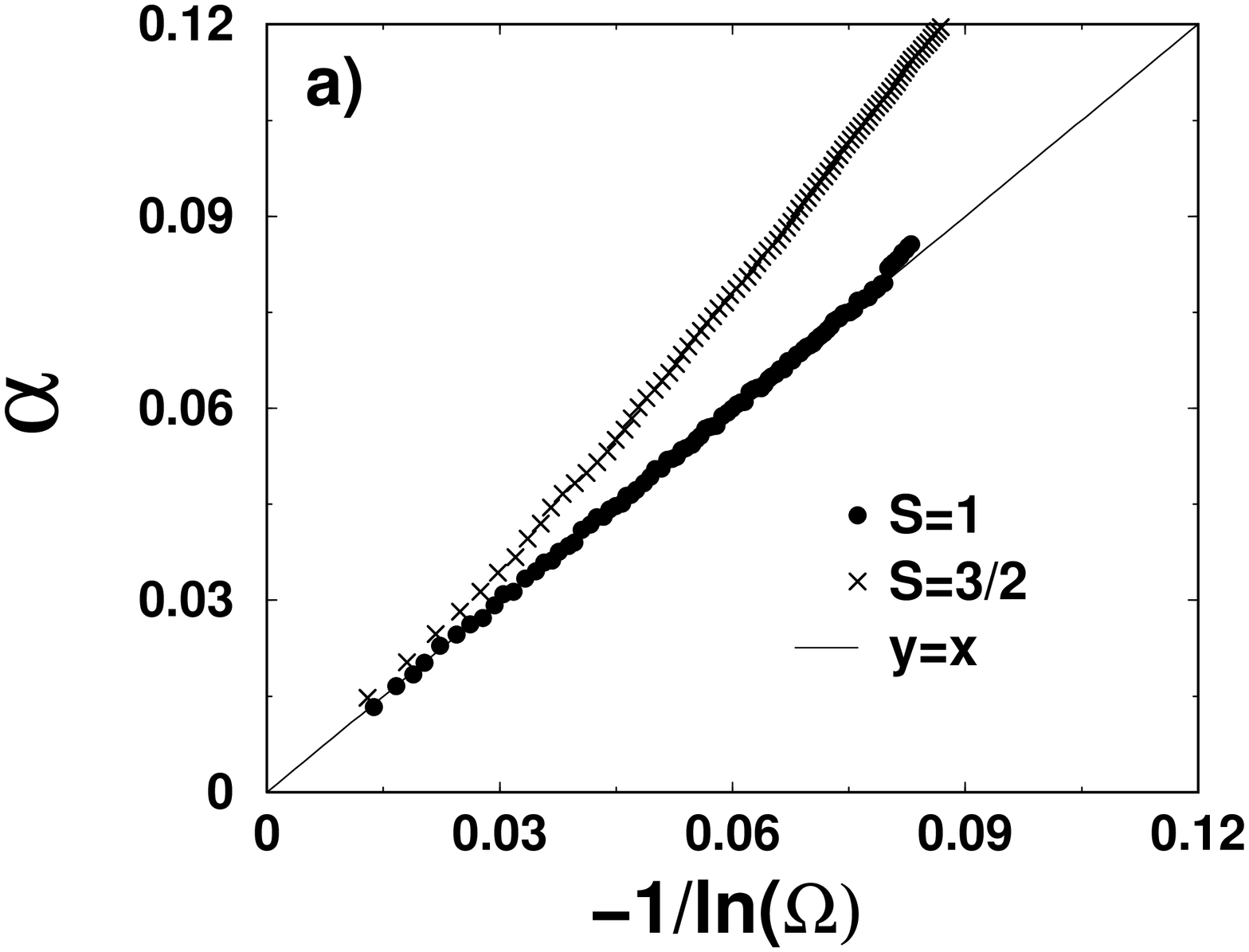}}
{\includegraphics[angle=-90,scale=0.175]{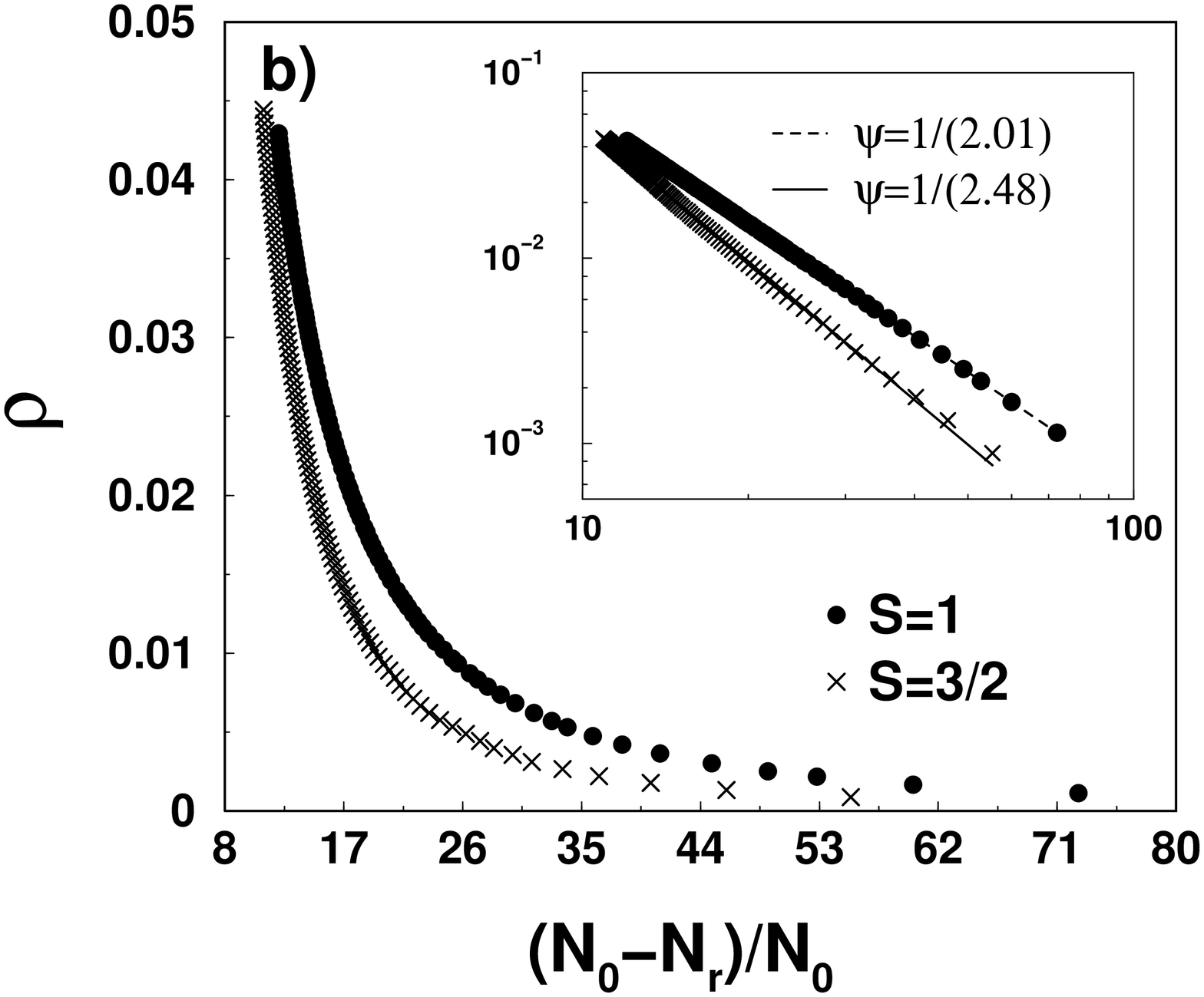}} \caption{\label{fig6} a) The exponent $\alpha$ describing the
asymptotic low energy behavior
 of the renormalized exchange distribution as a function of the scale dependent cut-off.
  b) Fraction of active spins as a function of the cut-off.
 For comparison we show the results for the spin-$1$ and spin-$3/2$ REHACs.
 In both cases the starting distribution is extremely disordered with $\delta=20$ (see text).}
\end{figure}

In order to check the existence of a random singlet phase in the
spin-$3/2$ chain we consider another class of distributions of
exchange couplings associated with {\em extreme}
disorder~\cite{igloi3,igloi4,mot}. These distributions are of the
form, $P(J) \propto J^{-1+1/\delta}$. For $\delta=1$ this reduces
to the gapless case of rectangular distributions considered
previously and for $\delta
> 1$, we have the extreme disordered cases. We now report our
results for the random spin-$3/2$ chain obtained with the modified
renormalization group procedure \cite{saguia}  for the case of an
extreme disordered distribution with $\delta=20$. In a random
singlet phase the fixed point distribution of interactions which
is attained when the cut-off $\Omega$ is sufficiently reduced,
takes the form
\begin{equation}
\label{2} P(J)=\frac{\alpha}{\Omega}\left(\frac{\Omega}{J}\right)^{1-\alpha}.
\end{equation}
The exponent $\alpha$ is a function of the cut-off $\Omega$  and varies as, $ \alpha=-1/ \ln \Omega$. Also for a
random singlet phase the fraction of remaining active spins $\rho$ as a function of the energy scale set by the
cut-off $\Omega$ \cite{fisher} is given by,
\begin{equation}
\rho=\frac{1}{L}=\frac{1}{|\ln\Omega|^{1/\psi}}. \label{4}
\end{equation}
The exponent $\psi$  establishes the connection between the
characteristic length $L$ and the energy scale $\Omega$. This is
an extension of the usual definition of a dynamic exponent
($\Omega^{-1} \propto \tau \propto L^{z}$) for the case of
logarithmic scaling~\cite{mucio}. In Fig.~\ref{fig6}a  we show the
exponents $\alpha$ obtained from the asymptotic form of the
exchange distributions after the cut-off $\Omega$ has been
sufficiently reduced. For comparison we show the results for a
spin-$1$ REHAC with the same original  extreme disorder
distribution for which a random singlet phase is clearly
established. We have considered here chains of size as large as
$L=4.5 \times 10^5$. In Fig.~\ref{fig6}b we show the density
$\rho=1/L$ of active spins as a function of the cut-off $\Omega$.
From this expression we extract the exponent $\psi$
(see~Eq.\ref{4}) which takes the value $\psi=1/(2.4)$ close to the
value $\psi=1/2$ expected for a random singlet phase
\cite{fisher}. As shown in this figure, for comparison, the
spin-$1$ chain has clearly converged to this phase within the same
scale of the cut-offs. Our results suggest that in this case of
extreme disorder, the spin-$3/2$ REHAC eventually reaches a random
singlet phase, although the convergence is very slow.

Recently another approach to the spin-$3/2$ chains has predicted the existence of a random singlet phase in these
chains, even for weak disorder \cite{refael,huse}. This is associated with spin-$1/2$ degrees of freedom. These
results are quite distinct from those obtained above where a random singlet phase is hardly evident even for
extremely disordered original distributions.

The decomposition of a chain of spins-$S$ in smaller spins, relies on projecting out the highest energy level of a
pair of spins-$S$. However the remaining excited states are kept in this procedure to maintain the correct number
of states. For example, in the case of a pair of spins-$1$ with a total of  nine states, the singlet ground state
and the first excited triplet are kept to yield the four states of the relevant antiferromagnetically coupled
spin-$1/2$ pair \cite{jolicoeur}.

The MDH elimination procedure can be generalized for finite temperatures and arbitrary
spins $S$. It is given by
\begin{equation}
\label{8}
J^{\prime}=\frac{2}{3}S(S+1)\frac{J_1J_2}{\Omega}W_S(\beta \Omega)
\end{equation}
where
\begin{equation}
\label{9} W_S(y)=\!\! \frac{(2S \!\!+1)^2-\!\! \sum_{i=0}^{i=2S}(2i \! +1)e^{-\frac{1}{2}i(i+1)y}\left[1 \!\!+
\!\! \frac{1}{2}i(i+\!\!1)\right]} {4S(S+1) \sum_{i=0}^{i=2S}(2i+1)e^{-\frac{1}{2}i(i+1)y}}
\end{equation}
Notice that for sufficiently high temperatures, the factor $\frac{2}{3}S(S+1)W_S(\beta \Omega)<1$ and the MDH
elimination procedure works in this case. A random singlet phase is reached in the sense that the asymptotic
distribution of exchange attains the form given by Eq.~\ref{2} at these temperatures. However, as $T$ is reduced
the problem becomes essentially non-perturbative, for spins $S \ge 1$, as the equation above generates coupling
larger than those eliminated. In particular at $T=0$ the excited states which reduce the factor $W_S$ from its
value $W_S(T=0)=1$ are now frozen. In fact {\em none} of the excited states play a role in the problem at zero
temperature. Notice that in our generalized renormalization scheme, degenerate perturbation theory is applied to
the {\em ground state} of the spin trio. We believe this is the main reason for the discrepancy between our
results and those obtained by the authors of Refs~\cite{refael,huse}. The consideration of excited states in the
problem favors the appearance of an infinite disorder random singlet phase, as occurs at finite temperatures.

\begin{figure}
 \includegraphics[angle=-90,scale=0.3]{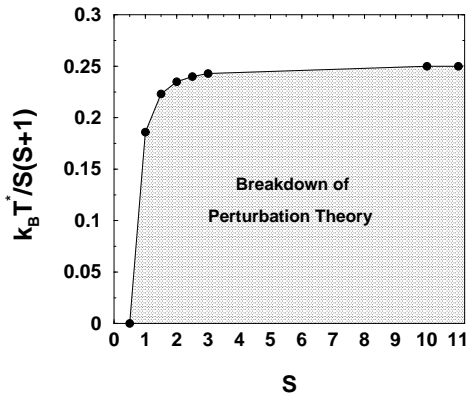}
 \caption{\label{fig8} Temperature $T^*$ below which the MDH perturbation theory breaks down
 for different values of the spin S. The energy $k_B T^*$ is in units of the cut-off $\Omega$ of the
 original exchange distribution.}
\end{figure}

In the limit $S \rightarrow \infty$ and $T \rightarrow 0$, replacing the sum by an integral and with a proper
renormalization of the Hamiltonian, Eqs.~\ref{8} and  \ref{9} yield $J^{\prime}=J_1J_2/4k_B T$, in agreement with
the result of Ref~\cite{MDH} for classical spins.  In Fig.~\ref{fig8} we show the temperature $T^*$ below which
the simple perturbative approach breaks down, for a given value of the spin of the random chain.

Notice that for rectangular distributions, the Griffiths phase of the spin-$1$ REHAC extends up to $G_c=0.45$ and
for spin-$3/2$ up to $G_c=0.11$. For random classical spin chains, the susceptibility $\chi \propto P(0)| \ln T| $
where $P(0)$ is the finite weight at the origin of the original distribution $P(J)$ \cite{MDH}. This weak
logarithmic singularity is similar to that expected for quantum chains at the border of the Griffiths phase
($Z\approx 1$) as if in this case, $G_c=0$.

We have studied a spin-$3/2$ REHAC using an extension of the
renormalization group procedure introduced by Ma, Dasgupta and Hu
\cite {MDH}. This method which considers larger clusters of spins
eliminates the difficulties associated with the perturbative
nature of the MDH procedure for the cases of spins $S \ge 1$. The
new procedure works very well for the spin-$3/2$ chain and never
bonds larger than those eliminated are generated for the
distributions used here. For rectangular distributions we found
the spin-$3/2$ REHAC presents a Griffiths phase up to a critical
value of disorder. We have also considered the case of extreme
disorder, where the starting exchange distributions are singular
for small values of the coupling. For values of the disorder
parameter as large as $\delta=20$ our results only suggest that a
random singlet phase will be asymptotically reached as the cut-off
of the distribution $\Omega \rightarrow 0$. We have compared our
results with those of another approach which predicts a random
singlet phase, associated with spin-$1/2$ degrees of freedom, even
for weak disorder. We attribute the difference between these
results and those we have obtained  to the fact that the former
approach takes into account excited states which mimic the effects
of temperature and favor the appearance of a random singlet phase.
Our results however are consistent  with those of Carlon {\it et
al.}~\cite{igloi} that find a random singlet phase in spin-$3/2$
chains for the case of extremely disordered distributions.

\begin{acknowledgments}

We would like to
thank Conselho Nacional de De\-sen\-vol\-vi\-men\-to Cient{\'{\i}}fico
e Tecnol\'ogico-CNPq-Brasil (PRONEX98/MCT-CNPq-0364.00/00), Fundac\~ao de Amparo
a Pesquisa do Estado do Rio de Janeiro-FAPERJ for partial financial
support.
\end{acknowledgments}

\end{document}